# Large scale Micro-Photometry for high resolution pH-characterization during electro-osmotic pumping and modular micro-swimming

Ran Niu*, Stanislav Khodorov, Julian Weber, Alexander Reinmüller, Thomas Palberg

*Institut für Physik, Johannes-Gutenberg Universität, Staudingerweg 7, 55128, Mainz*

**Abstract**

Micro-fluidic pumps as well as artificial micro-swimmers are conveniently realized exploiting phoretic solvent flows based on local gradients of temperature, electrolyte concentration or pH. We here present a facile micro-photometric method for monitoring pH gradients and demonstrate its performance and scope on different experimental situations including an electro-osmotic pump and modular micro-swimmers assembled from ion exchange resin beads and polystyrene colloids. In combination with the present microscope and DSLR camera our method offers a 2 µm spatial resolution at video frame rate over a field of view of 3920x2602 µm$^2$. Under optimal conditions we achieve a pH-resolution of 0.05 with about equal contributions from statistical and systematical uncertainties. Our quantitative micro-photometric characterization of pH gradients which develop in time and reach out several mm is anticipated to provide valuable input for reliable modeling and simulations of a large variety of complex flow situations involving pH-gradients including artificial micro-swimmers, microfluidic pumping or even electro-convection.

**Introduction**

Swimming at the micro-scale has attracted significant attention due to its potential applications as well as its inherent experimental and theoretical challenges. To overcome thermal fluctuations and viscous effects, nano-motors and micro-swimmers scavenge fuel from surrounding media, or harness power from external energy sources such as electric and magnetic fields, light, ultrasound, or combinations of these [1-3]. Special attention has been paid to phoretic pumps and swimmers, which rely on local field gradients (concentration gradient of dissolved species, acoustic gradient, temperature gradient etc.) for moving the adjacent solvent or realizing self-propulsion [2,4-6]. During the past decade, and under intense cross-fertilization by theoretical modeling and simulation [4,7-12] great experimental efforts were devoted to optimize and modify artificial micro-swimmers, e.g., for improved speed and direction control [13-18], or for implementation of specific functionalities, such as cargo transport [19-24], drug delivery [25], or sensor qualities [26-28]. Moreover, bioengineering applications benefited from advances in micro-fabrication [29]. Characterizations most often employ optical methods. Swimmer position, orientation, speed and acceleration are conveniently obtained from optical microscopy [1,3,30]. Holographic microscopy in addition was used for three-dimensional imaging of swimmer motion and orientation, as well as mapping of 3D solvent flow fields using buoyancy matched tracers [31,32]. An even more challenging task are detailed, precise and comprehensive measurements of (local) experimental boundary conditions. Farniya et al. managed to infer local diffusio-electric field strength in electro-osmotic pumping from the velocities of tracers with different electrophoretic mobilities [33]. The concentration distribution of electrolytes under conditions of inter-diffusion was determined by the fluorescence intensity of dyes added to the diffusing buffer [34]. A related technique was used to investigate diffusio-phoretic motion of passive colloids in specially designed cells [28]. Despite these advances, however, large scale measurements of fields or concentration gradients with high spatiotemporal resolution are still scarce. This in turn severely affects comparison of experimental observations to precise and detailed theoretical modeling [4,14,35-39].

In the present work, we address this problem for systems of phoretic pumps and micro-swimmers based on self-generated pH-gradients. Such pH-gradients are utilized by many artificial swimmers for propulsion and steering of collective behaviors. Examples include catalytic particles [4,23,40,41] or swimmers responding to an external pH-gradients, i.e. performing pH taxis [42,43]. Also for our micro-fluidic pumps and micro-swimmers, the central characteristic is a pH-gradient induced electro-osmotic (eo) flow along a charged surface. Such a gradient is here generated by an ion-exchange particle (IEX) exchanging trace amounts of residual cationic impurities for protons. Flows visualized in tracing experiments have been sucessfully compared to simulation and analytic theory [44]. However, different from most other approaches relying on a single active component, we take a modular approach to assemble different inactive components into active entities. Under certain experimentally tuneable flow conditions, a single *fixed* micro-pump can be ultilized to sort passive particles by size and assemble colloidal crystals [45,46]. Also a *mobile* IEX-type pump can assemble and couple passive colloidal particles. Here, however, breaking of the flow symmetry leads to propulsion of the whole self-assembled complex [47]. Such modular swimmers move at velocities of several µm/s over extended times. Swimming of complexes containing a regular arrangement of a few cargo particles is captured by a simple geometric model [30]. A wide range of further assembly types are formed by combinations of cationic IEX with other types of cargo or mixed type cargo (cationic IEX beads, anionic IEX beads, Janus swimmers, magnetic colloids etc.) including the possibility of passive or active "colloidal molecules" displaying long range attractive interactions based on the underlying eo flow and allowing for explorations of their intriguing self-assembly dynamics [48].

To introduce our novel characterization approach, we stick to the two most fundamental examples of IEX based micro-fluidics, i.e., ion exchange based micro-fluidic pumps and micro-swimmers with passive polystyrene cargo. In what follows, we present the photometric method for precise monitoring of the pH gradient over large scales covering the whole non-equilibrium environment of the active entities. We further show how to achieve this with high spatial and temporal resolution employing low cost commercial equipments and discuss the performance and scope of our novel approach.

## Method

Like in macroscopic pH measurements employing indicator paper or fluids, we here exploit the pH-dependent color change of suitable acid-base indicators (e. g. mixtures of universal indicators pH 0-5 and pH 4-10, Sigma Aldrich, Inc.). The samples are illuminated by white light and the transmitted light is analyzed for its color distribution. This is realized with a consumer DSLR (D700, Nikon, Japan) mounted on an optical microscope (DMIRBE, Leica, Germany) yielding a high spatio-temporal resolution. All automatic color processing and white balance were turned off. Combining a 5× magnification objective with a 0.63× mounting tube yields a field of view of 3920x2602 µm$^2$ at a resolution of 1.8 µm. The maximum frame rate is 5 fps in full frame (FX) single exposure or video frequency for full HD RGB images. The FX CMOS sensor is composed of 12.1x10$^6$ individual pixels each made sensitive for light in Red, Green, or Blue band by a suitable filter. Color discrimination on the sensor uses a Bayer pattern (Figure S1 in the Appendix). 14-bit raw format images are stored on a computer and converted using dcraw to 16-bit linear TIFF files before further processing. Separation of the three color channels (debayering) is accomplished using a home-written Python script, yielding three separate arrays for Red, Green and Blue channels.

According to the Beer-Lambert law, the absorbance $A$ of light transmitting through a sample at a given wavelength (or channel) is given as:

$$A = -\log_{10}(I/I_0) = \mu_\lambda\, c\, d, \qquad (1)$$

where $I_0$ and $I$ are the light transmitting through the reference (cell filled with filtered Milli-Q grade water) and the sample at the same illumination intensity; $\mu_\lambda$ is the wavelength dependent attenuation coefficient, $c$ is the concentration of the adsorbing agent and $d$ is the path length of light through the sample. Individual universal indicator fluids are optimized for color discrimination using human eye. However, these typically yield non-monotonous dependencies of $\mu_\lambda$ on pH. We therefore use a suitable indicator fluid mixture and calibrate the *ratio* of absorbance of two different channels (e.g. $A_{Blue}/A_{Red}$) to pH using buffer solutions of different pH. According to Eqn. (1), the absorbance ratio of two colors equals their ratio of attenuation coefficients, $\mu_{Blue}/\mu_{Red}$, which renders the calibration and measurements independent of indicator concentration and sample thickness. Making use of the reference cell further eliminates any dependence on illumination intensity

A typical measurement is shown in Figure 1a for buffer solutions in the pH range of (1.89-8.94) ± 0.02 with a step width of about 0.5. Noise reduction was performed by averaging over several frames. Debayering of averaged images yields three ratios of $\mu_\lambda$ at a fixed pH. Figure 1a shows that indeed the ratio of $\mu_{Blue}$ to $\mu_{Red}$ monotonically decreases with increasing pH. The statistical uncertainty of $\mu_{Blue}/\mu_{Red}$ is ± 0.015 as obtained from averaging 20 subsequent frames at a frame rate of 5 fps (thus dropping the effective temporal resolution to 4 s). As there is no functional form fitting to the data, we interpolated the measured data points for the calibration curve (pH versus $\mu_{Blue}/\mu_{Red}$) using linear interpolation as shown in Figure 1b.

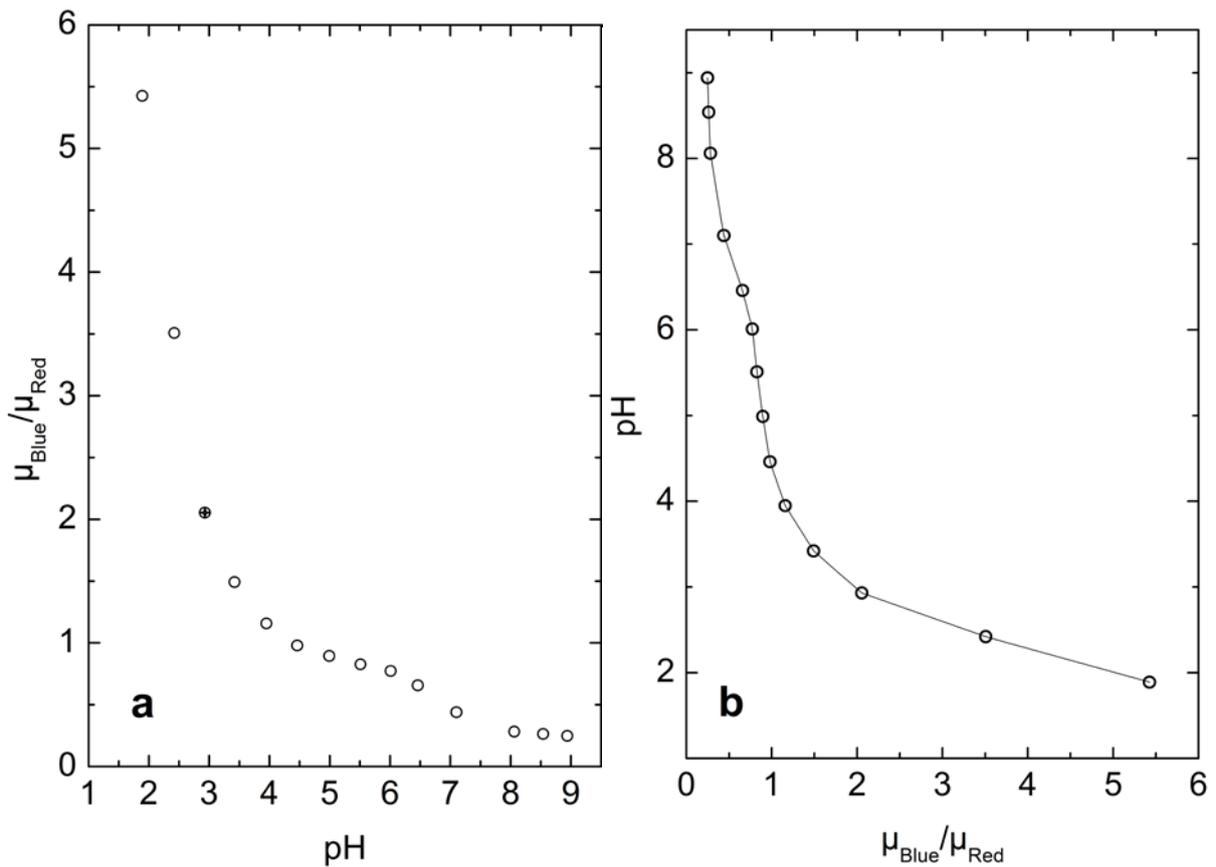

**Figure 1.** a) Adsorption coefficient ratio of blue to red $\mu_{Blue}/\mu_{Red}$ as a function of pH. The statistical error is below the symbol size (± 0.02 for the pH of buffer solution and ± 0.015 for $\mu_{Blue}/\mu_{Red}$ as indicated for the point at pH 3). b) Calibration curve: pH as a function of $\mu_{Blue}/\mu_{Red}$. The solid line interpolates the measured data points.

## Results

The performance of our photometric method was first tested on a stationary micro-fluidic pump consisting of an IEX bead of 45 μm in diameter (IEX45) fixed to a charged glass substrate at micro-molar residual salt concentration. The mechanism of this pump has been investigated in detail in our previous work [44]. Briefly, the IEX exchanges residual cationic impurities for stored protons and establishes a pH gradient which in turn induces diffusio-electric (*E*-) fields due to the difference in the cationic diffusion coefficients [49]. However, since there is no electric current, no electro-chemical reaction is possible, which could compromise pH measurements using acid-base indicators if happens [50]. The *E*-fields act on the electric double layer of the negatively charged substrate inducing an electro-osmotic (eo-) flow along the substrate. This flow converges at the IEX bead where it turns upward. This directed flow can be measured in tracer velocimetry experiments out to hundreds of microns until it becomes dominated by Brownian motion. Simulations, however, show that the flow reaches much further.

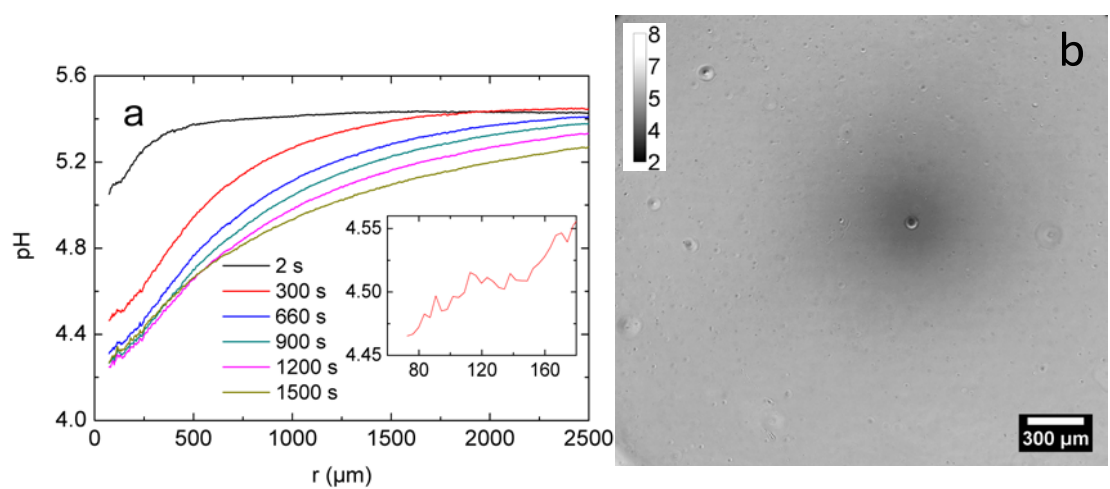

**Figure 2.** a) Radially averaged pH-values as a function of radial distance *r* from the center of IEX45 for different times after contact with water. Inset: an enlarged area of the curve at *t* = 300 s. b) pH map for an IEX45-based pump taken 5 min after the start of the experiment. pH-values as determined from the ratio $\mu_{Blue}/\mu_{Red}$ are shown using a non-linear grey chart as indicated.

Figure 2a quantitatively shows the temporal evolution of the pH gradient after an IEX45 contacts with water (image taking starts from 2 s). In the first 5 min, the pH-values in the immediate vicinity of IEX45 decrease steeply to ~4.4, and the pH gradient reaches out to a larger radial range. Figure 2b shows the stable pH gradient generated by an IEX45-based pump about 5 min after contact with water. The measurable gradient extends to a radial range *r* (measured from the center of the IEX) of about 600 μm. Due to the saturation with atmospheric $CO_2$, the initial background pH is about 5.5. The pH-value at *r* = 2500 μm decreases slowly over extended time scales as the gradient evolves (Figure 2a). In addition, as the ion exchange continues, the reaction equilibrium $CO_2(g) \leftrightarrow H_2CO_3 \leftrightarrow HCO_3^- + H^+$ shifts with decreasing salinity [51], which, however, does not influence the ion exchange process of the IEX as only protons are produced in this equilibrium. Note the excellent quality of the data obtained here by averaging over 20 subsequent images. This is highlighted also in the enlarged inset of Figure 2a where the maximum fluctuations are on the order of 0.02 pH units. Such good quality was found to be typical for optimum preparation conditions and stable pumping.

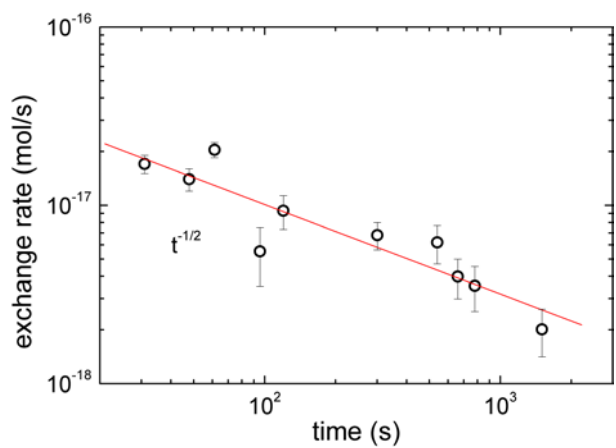

**Figure 3.** Exchange rate of IEX45 as a function of time averaged over 10 independent measurements. The red line indicates a power law decay proportional to $t^{-1/2}$.

The pH is defined as the negative decadic logarithm of the proton concentration. Integrating the proton concentration over the radial range of 2500 µm, we obtain the amount of exchanged protons at a given time $t$ after the start of the experiment. Taking the differences at different times, we can calculate the exchange rate of an individual bead of IEX45. Typical initial values for a micro-molar impurity concentration are on the order of $10^{-17}$ mol of protons exchanged per second. Figure 3 shows that this exchange rate decays with the square root of time, indicating that for a stationary IEX the exchange at low impurity concentrations is diffusion-limited [52].

The situation gets significantly more complex in the case of modular swimming. Now, the non-equilibrium and inhomogeneous environment challenges the high precision and resolution of the pH measurements. Figure 4 shows a time series of pH maps for a swimmer assembled from one IEX45 and a few PS20 cargo. The IEX particles were first placed in the cell at low areal density. Then one drop of a dilute suspension of cargo particles was added, which settled to the substrate in short time. By chance this IEX45 locked two PS20 at $t$ = 18 s. Note that the symmetry of the gradient is not immediately broken. At $t$ = 122.9 s, the swimmer has moved some 200 µm from its original position and its pH gradient becomes slightly asymmetric; meanwhile, another cargo was loaded (Figure 4b, taken from video 1 in the Appendix) changing the swimming direction and increasing swimming speed. Figure 4c (taken from video 2) taken after about 10 min shows the steady state pH gradient of this swimmer with three loaded PS20 moving at a speed of 3.8 µm/s. Note the steepening of the gradient in the front and the comet-trail like stretching in H+ range in the back of the swimmer. This shape of pH gradient is typical for swimmers with speeds above 1 µm/s and is formed irrespective of IEX and cargo size and cargo number when the swimmer has covered a sufficiently large distance as compared to the gradient width perpendicular to the motion direction (see videos 3 and 4 for examples of swimmers combining IEX67 with PS31).

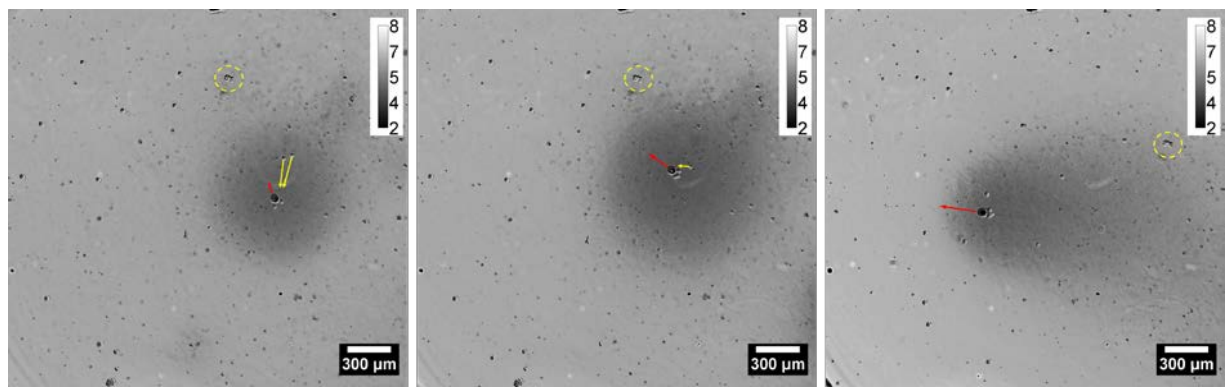

**Figure 4.** pH maps of a modular swimmer formed of a single IEX45 and PS20 cargo under swimming conditions (Image size 2701 x 2584 µm$^2$; scale as indicated). pH-values as determined from the ratio $\mu_{Blue}/\mu_{Red}$ are shown using a non-linear grey chart as indicated. The direction and magnitude of swimming motion is indicated by red arrows, the direction and magnitude of motion of PS20 are indicated by yellow arrows. a) IEX45 with two PS20 at $t$ = 18 s. The complex is moving very slowly in upward direction. Further cargos are on their way downward. The encircled dirt blob may give some orientation. b) Around $t$ = 122.9 s, a third PS20 gets loaded, the speed increases and the gradient of the swimmer becomes asymmetric. The swimmer now is too fast to load the second PS20 still approaching from the right. c) At $t$ = 600.9 s, after further cargo rearrangement, the modular swimmer with three assembled PS20 moves with a constant speed of 3.8 µm/s to the left. It has evolved a strongly asymmetric pH gradient trail which is dissolving slowly. As compared to the approximately symmetric situation at $t$ = 18 s, the low pH region is shrunk at the swimmer front and considerably extended in the back. Note that the right cargo particle is still following very slowly at a velocity of approximately 0.1 µm/s to the left.

By contrast to the stationary pump, a mobile modular swimmer thus can explore fresh regions of low H$^+$ concentration. The swimmers move straight as long as the cargo number does not change and the load is distributed in a symmetric fashion. Asymmetries and changes in the load distribution cause changes in the direction of motion [30]. In cells of large diameter (e.g. 20 mm), with continuous impurity supply (e.g. Na$^+$ leakage from the glass substrate), the ion-exchange rate and the resulting swimming speed and direction can therefore stay constant until the ion-exchange capacity is exhausted. Moreover, the pH distribution becomes stationary and its dependence on swimming speed can be investigated in detail. Although we show the shape of gradient and not its dependence on speed, the latter is an important quantity to be investigated.

With the cases shown in Figure 4, we also test the performance under non-ideal conditions. We here used an indicator solution originally mixed at higher concentration and left to age for a few hours. Due to some non-specified chemical reactions, one of the dyes precipitates in minor quantities. We checked that this did not disturb the pH detection. It however introduces "optical" impurities as can be readily seen comparing Figure 4 to Figure 2b. We therefore can test the influence of dirt on our method. A first result is immediately apparent from the smooth pH distribution in the vicinity of dirt or PS particles: both species are inactive regarding gradient and thus field generation. For quantitative measurements, we again average over 20 frames taken over 4 s starting at $t$ = 598.9 s. The swimmer speed of 3.8 µm/s introduces a total IEX displacement of 15.2 µm which is accounted for by averaging images with the IEX center taken as the origin. For asymmetric gradients, no radial averaging is possible. We therefore quantify the pH gradient using line analysis only for selected angles 0° ≤ Θ ≤ 180° with respect to the swimming direction (Θ = 0°). The results for the swimmer of Figure 4c (one IEX45 plus three PS20) moving at a constant speed of 3.8 µm/s are shown in Figure 5.

As compared to Figure 2a, the curves in Figure 5 are considerably more noisy. Several individual peaks stemming from precipitates and lone cargo particles distribute statistically along $r$ and are uncorrelated between different angles. Statistics could be improved much further by averaging over subsequent frames spaced by swimmer positions separated by more than the typical diameter of impurity Airy disk of about 1-5 µm. Alternatively, also averaging over small angular wedges of e.g. ± 2° can be employed. However, even without further measures, the data quality from this dirty case is already sufficient to obtain a good qualitative impression on the characteristic effects.

Within statistical uncertainty, the surface pH of 3.8±0.1 in the back and 3.9±0.1 in the front of the IEX are only marginally different. Gradients determined in the front of the swimmer (0° ≤ Θ ≤ 60°) are much steeper. The background pH is reached after about 400 µm. In the back of the swimmer (90° ≤ Θ ≤ 180°), the gradients get shallower extending out to some 1500 µm before vanishing. As already noted from the visual observation of the pH images this effect gradually increases with increasing Θ. Moreover, in the plateau or far-field region (1600 µm < r < 2500 µm) we note a strongly decreased background pH level in the IEX wake. The pH-values drop systematically with Θ increasing from 150° to 180°, i.e., the low pH trail dissolves only slowly.

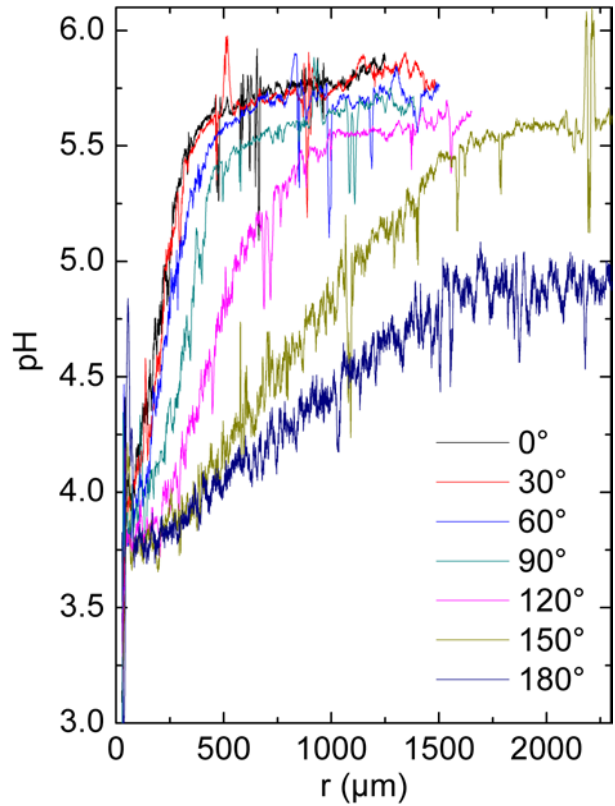

**Figure 5.** pH as a function of *r* measured along different angles Θ with respect to the moving direction. Data taken for 20 subsequent frames recorded starting at *t* = 598.9 s of the self-propelling swimmer shown in Figure 4c. Note the presence of narrow spikes signifying dirt particles or cargo particles.

**Discussion**

We have introduced a facile photometric method to monitor minute pH variations in the micro-environment of electro-osmotic pumps and modular micro-swimmers with high spatiotemporal resolution. For the present settings, we obtain a temporal resolution of 4 s within which we average over 20 images taken at a frame rate of 5 fps. The spatial resolution is 1.8 µm over a large field of view of 3920x2602 µm². However, both the spatial and temporal resolutions are tunable according to the application. The temporal resolution is set by the image recording speed, therefore faster swimmers may need a camera of larger frame rate to resolve the fast developing gradients carried along by the swimmer. The spatial resolution of the method is determined by the image quality (pixels per µm) set by the camera resolution and the magnification of the microscope. In the case of slow swimmers as investigated here, resolution may be sacrificed for statistical improvements. We explicitly note that the method was implemented with relatively inexpensive standard lab equipment (DSLR, low resolution microscope, standard desktop computer). Upgrades in spatial resolution can be realized by higher magnification objectives which can be combined with a motorized stage to maintain a large field of view. We think that for most applications video frame rate is sufficient. Use of fast RGB cameras with sufficient resolution probably leaves the low budget region.

One major advantage of our method is the independence of the pH signal on local indicator concentration and the cell thickness resulting from using the Beer-Lambert law and taking the ratio of $\mu_{Blue}$ to $\mu_{Red}$. This makes our measurements quite robust, and even strongly absorbing dirt particles yielded only a few manageable spikes in the radial pH distribution curves. The maximum achievable pH resolution for our method is mainly determined by statistical uncertainty of about 0.02 and the calibration of the indicator mixture. Averaging over the complete accessible pH range, the systematic uncertainty is about 0.015 for the particular mixture of universal indicators used here. As seen in Figure 1b in the pH range relevant for our IEX based micro-fluidic pumps and micro-swimmers (4 - 6.5) it is 0.08, however, an even better pH resolution is obtained at lower pH. Alternative dye mixtures are currently tested. The high accuracy of pH determination allowed the spatiotemporal characterization of the pH distribution around pumps and swimmers, and it thus enabled further analysis with assignment of kinetic laws. The optimal performance was, of course, obtained at a stationary pump (Figure 2) for which the radial symmetry allowed a large number of pixels for averaging. For micro-swimmers, the symmetry breaking only allows line averaging, thus there is increased noise (Figure 5). However, even without optimum preparation of samples, e.g. with dirt or precipitates of indicator as shown in Figure 4, our method still works well. Averaging over angular wedges or successive frames with correction for shifted origin may improve statistics. This approach seems most suitable for situations with stably locked passive cargo resulting in straight motion of constant speed. For active cargoes, e.g. catalytic Janus particles, their active motion as well as local reaction or flow field induced pH fluctuations may become a further challenge.

An important detail of our set-up is the low cell height of 0.5 mm. This assures a quick vertical homogenization of the pH such that layering effects are avoided. Layering effects are expected for large cell heights in which upward proton diffusion is slower than sideways convection. In fact, for the pumping and swimming conditions employed here, they are expected and found for cell heights significantly above 1 mm (e.g. 10 mm). While layering cannot be detected in transmission experiments, it is accompanied by a clear signature - a flow type change from $1/r$ to $1/r^2$ which can be conveniently checked by adding some tracer particles [44].

Utilizing our novel method, we have revealed a diffusion-limited ion exchange for IEX-based pumping and modular swimming at low impurity concentrations. For the modular micro-swimmers, we observe a time dependent pH gradient during pump to swimmer transition and a typical comet trail shaped pH gradient under stationary swimming conditions. A systematic investigation on the shape of gradient under the influence of varied experimental boundary conditions (e.g., substrate zeta potential, size of swimmer constituents, background salt concentration) remains to be done. Clearly, an asymmetric pH gradient will induce an angular variation of the diffusio-electric fields, and hence an angular dependence in the eo-flow. Its influence on the velocity and direction of modular micro-swimmers, as well as feed-back effects between the altered swimming and the trailing gradient, the so-called auto-chemotaxis [53], are interesting issues. In the present case of diffusion-limited ion exchange, straight propulsion into as yet undisturbed regions ensures a constant speed of modular swimmers assembled from one IEX and a constant number of passive cargo particles. This effect is more pronounced at larger cargo number since the velocities become larger. A pH distribution disturbed by other IEX particles encountered on the swimmer's way may, however, lead to interesting pH gradient mediated interactions. We further anticipate that curved swimming (initially caused e.g. by a lop-sided cargo distribution) may give rise to interesting situations, as soon as the trajectories overlap and the swimmer encounters the remains of its own trail. In this case one may expect a stabilization of circular swimming due to eo-inflow into the low pH region. Moreover, also for catalytic Janus cargoes, which themselves can alter the pH during $H_2O_2$ consumption, a different pH distribution may result in alternative motion patterns. Our photometric method seems excellently suited to explore such cases in great detail.

Moreover, our method can also be applied to measure the pH gradients generated by other phoretic swimmers, e.g., the large variety of catalytic Janus particles (insulator or bimetallic) of any shape (sphere, rod, nano-tree etc.) [1,2,4,7,54-56] or isotropic particles with built-in shape asymmetry (e.g., AgCl micro-stars [57]) or steered by light (e.g., negative auto-phototaxis [58]), all dispersed in their respective fuel solutions. To study a simple pump situation, such a particle could be fixed by optical tweezer or glue. Measured proton concentration gradients $\nabla c$ and fluxes $J(t)$ across the surface of a particle could first be used to test the predictions for different phoretic mechanisms (e.g. for catalytic Janus particles [59] or light sensitive $TiO_2$ particles [58]). Moreover, these parameters can be employed to calculate the local diffusio-electric potentials, and estimate fluid velocities as well as velocities of mobile versions of these active particles [4,49]. Both types of information would significantly enhance possibilities of further modeling. In principle, similar analysis may also apply for asymmetric particles in AC electric fields, where electro-osmosis and electro-hydrodynamics induce the assembly of asymmetric particles [60-62].

Using larger magnification objectives, our method should further allow sufficient spatial resolution to measure local pH gradients along the surfaces of active particles and micro-swimmers. This could be used to test the pH-taxis of flagellated bacteria and bio-hybrid micro-systems [63,64], or artificial swimmers [42,43]. Utilizing the photometric method, one could measure both the imposed global pH gradient and the local gradient generated by active particles. Again, quantified local gradients can be used to further estimate asymmetric reaction kinetics, imbalanced osmotic pressure across the particle surface [7,65] and explore their roles in pH taxis.

Finally, it may be interesting to apply our method to the complex collective behavior of groups of phoretic micro-swimmers [41,48,55,66]. It therefore could assist in clarifying experimentally the interactions among phoretic swimmers. These are nearly always very complex due to the interplay between gradients from individual particles and the perturbation from neighboring particles [4,5]. Different approaches of non-equilibrium statistical physics have been applied in studies on the self-assembly of active particles [3,67,68], however, only a few theories have explicitly taken the details of interactions into account [48,69]. We anticipate that mapping out the involved local gradients, their evolutions and fluctuations in sufficient detail will yield valuable complementary information to standard particle tracking approaches.

**Conclusion**

In this work, we present a facile photometric method to characterize the pH distribution and gradients in the microenvironment of active particles with high temporal and spatial resolution. We demonstrated the performance, scope and limits of the method by characterizing IEX based pumps with diffusion-limited ion exchange and modular swimmers during cargo uptake, acceleration and final stationary swimming states. We anticipate that our method will turn out to be very useful in characterizing pH gradient related issues like micro-swimming, pH- and chemo-taxis as well as collective behaviors. Moreover it will deliver important constraints to theoretical modeling and simulation. Our complementary approach will therefore also tighten the connection between experiment and theory.

**Acknowledgement**

It's a pleasure to thank Joost de Graaf, Christian Holm, Thomas Speck, Hartmut Löwen and Denis Botin for fruitful and stimulating discussions. We further thank Christopher Wittenberg for technical support in image processing and cross-checks of Python scripts and Jairo Orozco Sandoval for assistance in developing the method. Financial support of the DFG (SPP 1726, Grant No. Pa459/18-1, 2) is gratefully acknowledged.

**Appendix**

*Experimental*

The ion exchanger (IEX) spheres used for the gradient measurements were micro-gel based cationic IEX of diameters 20-100 µm (CGC50×8, Purolite Ltd, UK). They were manually sorted into diameter classes of 45 ± 1 µm and 67 ± 1 µm, denoted as IEX45 and IEX67, respectively. The model cargo particles are commercial, negatively charged polystyrene (PS) spheres stabilized by sulfate surface groups (MicroParticles GmbH, Germany) with diameters of 19.7 ± 0.2 µm and 31.1 ± 0.3 µm and electrophoretic mobilities of $\mu_{ep}$= -(2.6 ± 0.2)x$10^{-8}$ $m^2$ $V^{-1}$ $s^{-1}$ and $\mu_{ep}$= -(2.7 ± 0.3)x$10^{-8}$ $m^2$ $V^{-1}$ $s^{-1}$. Their lab codes are PS20 and PS31, respectively [30,44].

The pH gradient was measured using 1:5 (volume ratio) mixture of universal indicators (pH 0-5 and pH 4-10, Sigma Aldrich, Inc.). Before use, the solvent of the commercial indicators was exchanged for water by evaporating 98% of the alcohol at 60 °C followed by adding the same volume of deionized water.

The sample cell was constructed from a circular Perspex ring (inner diameter of 20 mm, height of 0.5 mm) fixed to microscopy slides (soda lime glass of hydrolytic class 3 by VWR international) by hydrolytically inert epoxy glue (UHU plus sofortfest, UHU GmbH, Germany) and dried for 24 h before use. Sample cells were washed with 1% alkaline solution (Hellmanex®III, Hellma Analytics) under sonication for 30 min, then rinsed with tap water and subsequently washed with doubly distilled water for several times. The zeta potential of the washed glass slides in deionized water is –(105 ± 5) mV [44] as determined by micro electro-phoresis.

Dye calibration was done by adding 0.18 mL of home-made buffer solutions of different pH ((1.89-8.94) ± 0.02, step width of ≈0.5) and a tiny amount (a few µL) of indicator solution into the sample cell.

For the pH measurements of IEX-based pumps, a single IEX45 was fixed to the glass slides by a tiny amount of epoxy glue (UHU plus sofortfest, UHU GmbH, Germany) and dried for 24 h. Afterward, 0.18 mL of deionized water and a few µL of indicator solution were added. For the gradient measurements of swimmers, a few IEX45 were placed inside the cell, followed by injection of 0.18 mL dilute deionized cargo suspension and a few µL of indicator solution. All the cells were quickly covered with another glass slide to avoid contamination by dust.

RGB color-images were taken using a consumer DSLR (D700, Nikon, Japan) mounted on an inverted scientific microscope (DMIRBE, Leica, Germany). The camera has a 12.1 megapixel FX-format CMOS sensor with image dimensions of 4256x2832. Using a 5x magnification objective and a LM-Scope tube of 0.63x magnification connected between camera and the microscope yields a black and white spatial resolution of 1.8 µm and a field of view of 3920x2602 µm². From these images, both the pH gradient and the velocity of swimmers or tracer particles were obtained using home-written Python scripts.

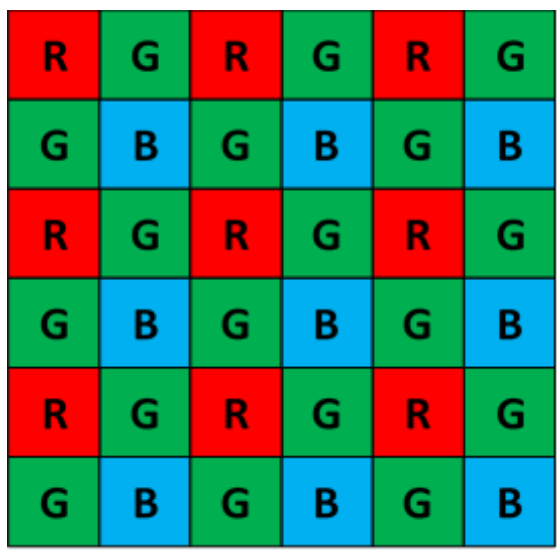

**Figure S1.** A schematic Bayer pattern of the color filter above a CCD camera.

*Videos:*

**Video 1:** Evolution of pH gradient with time from two PS20 assembling at one IEX45 to the formation of a swimmer. The gradient becomes asymmetric after 122.6 s. 45x real time speed.

**Video 2:** pH gradient of the modular swimmer formed from one IEX45 and three PS20 cargos at a speed of 3.8 µm/s. 60x real time speed.

**Video 3:** pH gradient of the modular swimmer formed from one IEX67 and one PS31 at a speed of 1.1 µm/s. 90x real time speed.

**Video 4:** pH gradient of the modular swimmer formed from one IEX67 and four PS31 at a speed of 1.9 µm/s. 90x real time speed.